\documentclass[12pt,a4paper]{article}
\usepackage{a4wide}
\usepackage{latexsym}
\usepackage{epsf}
\usepackage{amssymb}
\def\makeatletter{\catcode`\@=11}
\makeatletter
\def\mathbox#1{\hbox{$\m@th#1$}}%
\def\math@ccstyles#1#2#3#4#5#6#7{{\leavevmode
      \setbox0\mathbox{#6#7}%
      \setbox2\mathbox{#4#5}%
      \dimen@ #3%
      \baselineskip\z@\lineskiplimit#1\lineskip\z@
      \vbox{\ialign{##\crcr
             \hfil \kern #2\box2 \hfil\crcr
             \noalign{\kern\dimen@}%
             \hfil\box0\hfil\crcr}}}}
\def\mathaccstyles{\math@ccstyles\maxdimen}
\def\maththroughstyles{\math@ccstyles{-\maxdimen}}
\def\unitmatrixDT%
 {\maththroughstyles{.45\ht0}\z@\displaystyle {\mathchar"006C}\displaystyle 1}
\def\tfrac#1#2{{\textstyle{#1\over #2}}}

\makeatletter
\@addtoreset{equation}{section}
\makeatother

 \def\unit{\hbox to 3.3pt{\hskip1.3pt \vrule height 7pt width .4pt \hskip.7pt
\vrule height 7.85pt width .4pt \kern-2.4pt
\hrulefill \kern-3pt
\raise 4pt\hbox{\char'40}}}

\begin{document}

\begin{flushright}
\small
UG/16--98\\
IFT-UAM/CSIC-98-27\\
{\bf hep-th/9901055}\\
January $13$th, $1999$
\normalsize
\end{flushright}

\begin{center}


\vspace{.7cm}

{\Large {\bf The Super D9-Brane and its Truncations}}

\vspace{.7cm}


{\bf\large Eric Bergshoeff}${}^{\diamondsuit}$
\footnote{E-mail: {\tt E.Bergshoeff@phys.rug.nl}},
{\bf\large Mees de Roo}${}^{\diamondsuit}$
\footnote{E-mail: {\tt M.de.Roo@phys.rug.nl}},
{\bf\large Bert Janssen}${}^{\spadesuit\clubsuit}$
\footnote{E-mail: {\tt Bert.Janssen@uam.es}},\\
{\bf\large and Tom\'as Ort\'{\i}n}${}^{\spadesuit\clubsuit}$
\footnote{E-mail: {\tt tomas@leonidas.imaff.csic.es}}
\vskip 0.4truecm

${}^{\diamondsuit}$\ {\it Institute for Theoretical Physics, Nijenborgh 4,
9747 AG Groningen\\ The Netherlands}

\vskip 0.2cm
${}^{\spadesuit}$\ {\it Instituto de F\'{\i}sica Te\'orica, C-XVI,
Universidad Aut\'onoma de Madrid \\
E-28049-Madrid, Spain}

\vskip 0.2cm

${}^{\clubsuit}$\ {\it I.M.A.F.F., C.S.I.C., Calle de Serrano 113 bis\\ 
E-28006-Madrid, Spain}

\vspace{.7cm}


{\bf Abstract}

\end{center}

\begin{quotation}

\small

We consider two inequivalent truncations of the super D9--brane: the
``Heterotic'' and the ``Type I'' truncation.  Both of them lead to an
$N=1$ nonlinear supersymmetrization of the $D=10$ cosmological constant.
The propagating degrees of freedom in the Heterotic and Type I
truncation are given by the components of a $D=10$ vector multiplet and
a single Majorana-Weyl spinor, respectively.  
As a by-product we find that, after the Type I truncation, the Ramond-Ramond
super ten-form provides an
interesting reformulation of the Volkov-Akulov action. These results can
be extended to all dimensions in which spacetime filling D-branes exist,
i.e. $D=3,4,6$ and $10$.  

\end{quotation}

\newpage

\pagestyle{plain}

\section*{Introduction}

It is well-known that the dynamics of Dp--branes \cite{Po} is
described by a Dirac-Born-Infeld (DBI) action \cite{Le}.  For super
Dp--branes the kappa--symmetric generalization of the DBI action has
been given in \cite{St3,St,kn:BT,Sl1,Sl2}\footnote{Super Dp-brane
  equations of motion with simultaneous world-volume and space-time
  supersymmetry have been constructed in \cite{emb}.}.  The case $p=9$
corresponds to a spacetime filling brane, the super D9--brane.  It
leads to a supersymmetrization of the $D=10$ Born-Infeld (BI) action
in a IIB supergravity background \cite{kn:BT}. For a flat background
the explicit supersymmetry rules, after gauge-fixing the worldvolume
reparametrizations and the kappa--symmetry, have been given in
\cite{Sl2}. The super D9--brane has also been studied in \cite{sd9}.
Spacetime filling branes in $D=6$ have recently been discussed in
\cite{Ketov}.

The D9--branes play an important role in the description of the Type I
$SO(32)$ superstring in the sense that they provide the Chan-Paton
factors of the open superstring. In this description a truncation is
performed using the worldsheet parity operator $\Omega$ \cite{Po2}.
This leads to a specific ``Type I'' truncation of the IIB supergravity
background. In the S--dual version the D9-branes are replaced by NS--9B
branes, the worldsheet parity operator $\Omega$ by the fermion number
operator $(-)^{F_L}$ and the Type I $SO(32)$ superstring by the
Heterotic $SO(32)$ Superstring \cite{Hull,Hull1}.  The truncation using
$(-)^{F_L}$ leads to a so-called ``Heterotic'' truncation of IIB
supergravity. Both the Type I and Heterotic truncations preserve a
linear $N=1$ supersymmetry of the supergravity fields.

It is the purpose of this letter to extend the Heterotic and Type I
truncations of the IIB supergravity background to the action for a
single super D9--brane \footnote{ Only the Type I truncation of D9 (as
well as the Heterotic truncation of NS--9B) have a natural origin in
string theory. It would be interesting to see whether the Heterotic
truncation of D9 (and the Type I truncation of NS--9B) also have a
natural place in string theory.}.  We find that in both cases the
surviving $N=1$ supersymmetry is nonlinearly realized on the surviving
worldvolume fields. For the Heterotic (Type I) truncation these
worldvolume fields are given by the components of an Abelian $D=10$
vector multiplet (a single $D=10$ Majorana-Weyl fermion). The
organization of this letter is as follows.  In Section 1 we will first
introduce a formulation of IIB supergravity in the string frame that
will be needed to describe the super D9-brane action.  A new feature is
that we have added two 10--forms to the IIB supergravity multiplet on
which the IIB supersymmetry algebra is realized. There existence was
already suggested in \cite{Hull}.  These 10--forms, which are also
employed in \cite{recent}, are needed to write down a gauge-invariant
Wess--Zumino (WZ) term for the D9--brane (NS--9B brane).  We will give
the $\mathbb{Z}_2$ symmetries that lead to the Heterotic and Type I
truncation of the IIB supergravity multiplet. For later use we also
include the flat background truncation, which actually preserves an
$N=2$ global supersymmetry.  In Section 2 we introduce the super
D9--brane and extend the $\mathbb{Z}_2$ symmetries of the IIB
supergravity background to the full super D9--brane action. These
$\mathbb{Z}_2$ symmetries define the Heterotic and Type I truncations of
the super D9--brane that will be discussed in Section 3. Finally, our
Conclusions can be found in Section 4.


\section{IIB Supergravity and its Truncations}

Our starting point is the IIB supergravity multiplet whose supersymmetry
rules, in the string frame, are given by\footnote{We ignore terms
bilinear or higher order in the fermions.}

\begin{equation}
\label{eq:susiibstringgeneral}
\begin{array}{rcl}
\delta e_{\mu}{}^{a} & = & 
\bar{\epsilon}\Gamma^{a} \psi_{\mu}\, ,\\
& & \\
\delta \psi_{\mu} & = & 
{\cal D}_\mu \epsilon 
-\frac{1}{8}\not\!\! {\cal H}_{\mu}\sigma^3 \epsilon 
+\frac{1}{16}e^{\varphi} \sum_{n=1}^6
\frac{1}{(2n-1)!} \not\!G^{(2n-1)} 
\Gamma_\mu {\cal P}_{n}\epsilon\, , \\
& & \\
\delta {\cal B}_{\mu\nu} & = & 
2\bar\epsilon \sigma^{3} \Gamma_{[\mu}  \psi_{\nu]}\, , \\
& & \\
\delta {\cal B}^{(10)}{}_{\mu_{1}\ldots \mu_{10}} 
&=&  e^{-2\varphi}\bar{\epsilon} \sigma^3\left(
 10  \Gamma_{[\mu_{1}\ldots \mu_{9}} 
\psi_{\mu_{10}]} - \Gamma_{\mu_{1}\ldots \mu_{10}} \lambda \right) \, ,\\
& & \\
\delta C^{(2n-2)}{}_{\mu_{1}\ldots\mu_{2n-2}} & = &
-(2n-2)  e^{-\varphi} \bar\epsilon {\cal P}_{n} 
\Gamma_{[\mu_{1}\ldots\mu_{2n-3}} 
\left(\psi_{\mu_{2n-2}]} -\frac{1}{2(2n-2)}\Gamma_{\mu_{2n-2}]}\lambda
\right)\\
& & \\
& &  + \frac{1}{2}(2n-2)(2n-3) C^{(2n-4)}{}_{[\mu_{ 1}\ldots\mu_{2n-4}}
\delta {\cal B}_{\mu_{2n-3}\mu_{2n-2}]}\, ,\\
& & \\
\delta\lambda & = & \left(\not\!\partial\varphi
-\frac{1}{12}  \not\!\! {\cal H}\sigma^{3}\right) \epsilon
+\frac{1}{4}e^{\varphi} \sum_{n=1}^6 
\frac{(n-3)}{(2n-1)!}\not\! G^{(2n-1)} 
{\cal P}_{n}\epsilon\, , \\
& & \\
\delta\varphi & = & \frac{1}{2}\bar\epsilon\lambda\, ,\\
\end{array}
\end{equation}

\noindent where

\begin{equation}
\label{Pn}
  \begin{array}{rcl}
{\cal P}_{n} & = & 
\left\{
\begin{array}{l}
\sigma^{1}\, ,\hspace{.5cm} n\,\, {\rm even}\, \\
\\
i\sigma^{2}\, ,\hspace{.5cm} n\,\, {\rm odd}\, \\
\end{array}
\right. \\
\end{array}
\end{equation}

\noindent The IIB field-strengths are defined in form language by
\cite{kn:GHT}

\begin{equation}
\begin{array}{rcl}
{\cal H} & = & 3d{\cal B}\, ,\\
& & \\
G^{(2n+1)} & = & dC^{(2n)} -{\cal H}C^{(2n-2)}\, .\\
\end{array}
\end{equation}

\noindent These curvatures are invariant under the gauge transformations

\begin{equation}
\begin{array}{rcl}
\delta_{\Sigma} {\cal B} & = & d\Sigma_{(NS)}\, ,\\
& & \\
\delta_{\Sigma} C^{(2n)} & = & d\Sigma_{(RR)}^{(2n-1)} -
  \Sigma_{(RR)}^{(2n-3)}{\cal H}\, .\\
\end{array}
\end{equation}

\noindent In component language the curvatures are given by

\begin{equation}
\begin{array}{rcl}
{\cal H} & = & 3\partial {\cal B}\, ,\\
& & \\
G^{(2n-1)} & = & (2n-1)\left\{ \partial C^{(2n-2)} 
-{\textstyle\frac{1}{2}}(2n-2)(2n-3)
\partial{\cal B} C^{(2n-4)}\right\}\, ,\\
\end{array}
\end{equation}

\noindent where all indices (not shown explicitly) are assumed to be
completely antisymmetrized with weight one. The RR potentials
$C^{(2n-2)}$ are independent for $n=1,2,3$.  The potentials for $n=4,5$
are related to those of $n=2,1$ via the relations

\begin{equation}
G^{(7)} \equiv - {}^\star G^{(3)}\, , 
G^{(9)} \equiv {}^\star G^{(1)}\,,
\end{equation}

\noindent where ${}^* G$ is the Hodge dual.  The curvature $G^{(5)}$ is
selfdual. 

Note that we have introduced two 10--forms ${\cal B}^{(10)}$ and
$C^{(10)}$.  Their 11-form curvatures are identically zero and they do
not describe any dynamical degree of freedom. They are needed in order
to write down a gauge-invariant and supersymmetric WZ term in the action
of a super D9--brane (NS--9B brane). 

Note that IIB supergravity is {\sl not} invariant under a target-space
parity transformation, which takes the selfdual, chiral IIB supergravity
into the anti-selfdual anti-chiral IIB supergravity.  Under target-space
parity transformations, only $C^{(0)}$, $C^{(2)}$ and $C^{(4)}$ are
pseudotensors and get an extra minus sign while the remaining RR
potentials, which are defined via Hodge duality, are tensors and do not
get any extra signs. 

The IIB supersymmetry rules (\ref{eq:susiibstringgeneral}) realize
the following IIB supersymmetry algebra (bosonic symmetries only)

\begin{equation}
\begin{array}{rcl}
[\delta(\epsilon_1), \delta (\epsilon_2)] & = & \delta_{{\rm gct}}(a^\mu)
 + \delta(\Sigma_{(RR)}) + \delta(\Sigma_{(NS)}) ,\\
\end{array}
\end{equation}

\noindent where the transformation parameters on the right-hand 
side are given by

\begin{equation}
\begin{array}{rcl}
  a^\mu & = & \bar\epsilon_2\Gamma^\mu\epsilon_1\,,\\
  &&\\
  \Sigma_{(RR)}^{(2n-1)}{}_{\mu_1\ldots\mu_{2n-1}} & = & 
a^\rho C^{(2n)}_{\mu_1\ldots\mu_{2n-1}\rho}
      + e^{-\varphi}\bar\epsilon_2{\cal P}_n \Gamma _{\mu_1\ldots\mu_{2n-1}}
       \epsilon_1  \\
  &&\\
  &&\qquad -(2n-1)C^{(2n-2)}_{[\mu_1\ldots\mu_{2n-2}}
    \bar\epsilon_2\sigma^3 \Gamma_{\mu_{2n-1}]}\epsilon_1\,,   \\
  &&\\
  (\Sigma_{(NS)})_{\mu} &=& a^\rho {\cal B}_{\mu\rho}
                - \bar\epsilon_2\sigma^3\Gamma_{\mu}\epsilon_1\,,\\
  &&\\
  (\Sigma_{(NS)})_{\mu_1\ldots\mu_9} &=& a^\rho 
      {\cal B}^{(10)}{}_{\mu_1\ldots\mu_9\rho}
         - e^{-2\varphi}
\bar\epsilon_2\sigma^3\Gamma_{\mu_1\ldots\mu_{9}}\epsilon_1\, .
\end{array}
\end{equation}

\noindent Note that on the ten-form potentials the general coordinate
transformation cancels against a field-dependent gauge transformation.
 
 There are five $\mathbb{Z}_{2}$ symmetries of the IIB supermultiplet
 \footnote{These symmetries correspond to involutions of the IIB
   superalgebra \cite{kn:T}.}.  The first two are called
 ``$\sigma^3$'' symmetries and are given by

\begin{equation}
\label{s3}
\left\{
\begin{array}{rcl}
f  & \rightarrow & \pm\sigma^{3}f\, ,     \\
& & \\
C^{(2n-2)} & \rightarrow & -C^{(2n-2)}\, , \hskip .5truecm \forall n\, ,\\
\end{array}
\right.
\end{equation}

\noindent where $f$ stands for any fermion doublet.  The next two
$\mathbb{Z}_2$ symmetries are called ``$\sigma^1$'' symmetries and read:

\begin{equation}
\label{s1}
\left\{
\begin{array}{rcl}
f  & \rightarrow & \pm\sigma^{1}f\, ,     \\
& & \\
C^{(2n-2)} & \rightarrow & (-1)^{n} C^{(2n-2)}\, ,\\
\end{array}
\right.
\hspace{1cm}
\left\{
\begin{array}{rcl}
{\cal B} & \rightarrow -{\cal B}\, ,\\
& & \\
{\cal B}^{(10)} & \rightarrow -{\cal B}^{(10)}\, .\\
\end{array}
\right.
\end{equation}

\noindent Finally, the fifth discrete symmetry is just the
$\mathbb{Z}_{2}$ grading of the $N=2B$ superspace which reverses the
sign of all fermions. It is associated to a truncation from $N=2B$ to
$N=0$ supersymmetry in which only the bosonic fields remain. We will not
discuss this grading any further. 
 
The $\mathbb{Z}_2$ symmetries lead to truncations that are defined by
setting to zero all fields that change sign under the corresponding
$\mathbb{Z}_2$. The surviving supersymmetry is determined by the
requirement that the fields that have been set equal to zero remain zero
under supersymmetry.  This leads to two possibilities.  The first is, in
both the $\sigma^3$ and the $\sigma^1$ cases, to keep local $N=1$
supersymmetry. The second possibility is to keep two {\rm global}
supersymmetries, but to set more fields than those that change sign
under the $\mathbb{Z}_2$ transformation, equal to zero. In fact, in this
last case only a flat zehnbein remains and supersymmetry is realized in
a trivial way. Nevertheless, this flat case is relevant later when the
IIB multiplet appears as a background for the D9-brane. 

The truncations with local $N=1$ supersymmetry will be referred to as
the ``Heterotic'' truncations $(\pm\sigma^3)$, and the ``Type I''
truncations $(\pm\sigma^1)$. We summarize the truncations below. 

\noindent{\bf Flat Background truncation:}
\begin{equation}
e_\mu{}^a  =  \delta_\mu{}^a\, ,\hskip .5truecm ({\rm all\ other\ fields\
 zero})\, .
\end{equation}

\noindent There are two unbroken supersymmetries with {\sl constant}
parameter $\epsilon$. Note that for simplicity we have set the dilaton
$\varphi$ equal to zero. 

\noindent{\bf Heterotic truncations:}

\begin{equation}
\left\{
\begin{array}{rcl}
C^{(2n-2)} & = & 0\, ,
\hspace{.5cm} n=1,\ldots,6\, ,  \\
& & \\
(1\mp\sigma^{3})f & = & 0\, . \\
\end{array}
\right.
\end{equation}

\noindent The transformation rules 
of the resulting $N=1$ Heterotic supergravity theory are:

\begin{equation}
\label{heterotic}
\begin{array}{rcl}
\delta e_{\mu}{}^{a} & = & 
\bar{\epsilon}\Gamma^{a} \psi_{\mu}\, ,\\
& & \\
\delta \psi_{\mu} & = & 
{\cal D}_\mu \epsilon 
\mp\frac{1}{8}\not\!\! {\cal H}_{\mu} \epsilon\, , \\
& & \\
\delta {\cal B}_{\mu\nu} & = & 
\pm 2\bar\epsilon \Gamma_{[\mu}  \psi_{\nu]}\, , \\
& & \\
\delta {\cal B}^{(10)}{}_{\mu_{1}\ldots \mu_{10}} 
&=& \pm  \bar{\epsilon}\Gamma_{[\mu_{1}\ldots \mu_{9}} 
\psi_{\mu_{10}]}\, ,\\
& & \\
\delta\lambda & = & \left(\not\!\partial\varphi
\mp\frac{1}{12}  \not\!\! {\cal H}\right) \epsilon\, , \\
& & \\
\delta\varphi & = & \frac{1}{2}\bar\epsilon\lambda\, .\\
\end{array}
\end{equation}

The unbroken supersymmetry parameter satisfies $\epsilon =
\pm\sigma^3\epsilon$. 

These two multiplets are associated to the effective field theory of
the Heterotic string based on right--handed and left--handed
worldsheet fields, respectively.  They are related by the duality
transformation ${\cal B}\rightarrow -{\cal B}$, ${\cal B}^{(10)}
\rightarrow -{\cal B}^{(10)}$.  We observe that in the Heterotic frame
${\cal H}$ occurs as torsion, i.e., it can be combined with the
spin-connection in ${\cal D}_\mu \epsilon$. As in the IIB supergravity
theory, the Heterotic supergravity multiplet is not invariant under
target--space parity transformations which interchange the chirality
of spinors.

\noindent{\bf Type~I truncations:}

\begin{equation}
\left\{
\begin{array}{rcl}
C^{(2n-2)} & = & 0\, ,
\hspace{.5cm} n=1,3,5\, ,  \\
& & \\
{\cal B} & = & 0\, ,\\
& & \\
{\cal B}^{(10)} & = & 0\, ,\\
& & \\
(1\mp\sigma^{1})f & = &0\, . \\
\end{array}
\right.
\end{equation}

\noindent The transformation rules  are:

\begin{equation}
\label{typeI}
\begin{array}{rcl}
\delta e_{\mu}{}^{a} & = & 
\bar{\epsilon}\Gamma^{a} \psi_{\mu}\, ,\\
& & \\
\delta \psi_{\mu} & = & 
{\cal D}_\mu \epsilon 
\pm\frac{1}{8\cdot 3!}e^{\varphi}  \not\!G^{(3)} 
\Gamma_\mu \epsilon\, , \\
& & \\
\delta C^{(2)}{}_{\mu\nu} & = &
\mp 2 e^{-\varphi} \bar\epsilon \Gamma_{[\mu} 
\left(\psi_{\nu]} -\frac{1}{4}\Gamma_{\nu]}\lambda
\right)\, ,\\
& & \\
\delta C^{(10)}{}_{\mu_1\ldots \mu_{10}} &=& 
\mp  10 e^{-\varphi} \bar\epsilon 
\Gamma_{[\mu_{1}\ldots\mu_9} 
\left(\psi_{\mu_{10}]} -\frac{1}{20}\Gamma_{\mu_{10}]}\lambda
\right)\, ,\\
& & \\
\delta\lambda & = & \left(\not\!\partial\varphi 
\mp \frac{1}{2\cdot 3!}e^{\varphi} \not\! G^{(3)} \right)\epsilon\, , \\
& & \\
\delta\varphi & = & \frac{1}{2}\bar\epsilon\lambda\, ,\\
\end{array}
\end{equation}

\noindent where the unbroken supersymmetry parameter satisfies
$\epsilon = \pm\sigma^1\epsilon$.

These two $N=1$ supergravity multiplets are associated to the effective
field theories of Type~I strings.  Observe that in the Type I frame
$G^{(3)}=3\partial C^{(2)}$ does {\sl not} occur as torsion. These two
$N=1$ theories are related by the duality transformation
$C^{(2)}\rightarrow -C^{(2)}$, $C^{(10)} \rightarrow - C^{(10)}$.


\section{The Super D9--Brane}

In this Section we introduce the action for the super D9--brane in
curved IIB superspace.  Using the notation of Ref.~\cite{kn:BT}, the
worldvolume fields are the supercoordinates and the Born-Infeld vector

\begin{equation}
\{Z^{M},V_{i}\}\, ,
\hspace{1cm}
Z^{M}=(x^{\mu},\theta^{\dot{\alpha} I})\, ,
\end{equation}

\noindent  where $\mu=0,\ldots,9$, $i=0,\ldots,9$,
$\alpha=1,\ldots,32$, $I=1,2$ and the string-frame background
superfields are

\begin{equation}
\{\varphi\, ,\ E_{M}{}^{A}\, ,\ {\cal B}_{MN}\, ,\ 
{\cal B}^{(10)}_{M_1\ldots M_{10}}\, ,\
C^{(2n-2)}{}_{M_{1}\ldots M_{2n-2}}\}\, ,
\hspace{1cm}
n=1,\ldots,6\, .
\end{equation}

\noindent The action is given by

\begin{equation}
\label{ad9}
S^{\rm (D9)} = S_{\rm DBI}^{\rm (D9)} + S_{\rm WZ}^{\rm (D9)}\, ,
\end{equation}

\noindent with Dirac-Born-Infeld (DBI) action

\begin{equation}
\label{D9BI}
S_{\rm DBI}^{\rm (D9)} = -\int_{M^{10}} d^{10}\xi e^{-\varphi} 
\sqrt{|g_{ij}+{\cal F}_{ij}|}\, ,
\end{equation}

\noindent and Wess-Zumino (WZ) term

\begin{equation}
\label{D9WZ}
S_{\rm WZ}^{\rm (D9)}= \int_{M^{10}} C e^{{\cal F}}\, ,
\end{equation}

\noindent where we have set the D9-tension equal to one.  The tensor
${\cal F}$ is given by

\begin{equation}
{\cal F}_{ij}= F_{ij}-{\cal B}_{ij}\, ,
\hspace{1cm}
F_{ij}=2\partial_{[i} V_{j]}\, , 
\hspace{1cm}
{\cal B}_{ij}= \partial_{i} Z^{M}\partial_{j} Z^{N}{\cal B}_{NM}\, .
\end{equation}

\noindent Furthermore

\begin{equation}
g_{ij} = E_{i}{}^{a} E_{j}{}^{b}\eta_{ab}\, ,  
\hspace{1cm}
E_{i}{}^{A}=\partial_{i}Z^{M}E_{M}{}^{A}\, ,
\hspace{1cm}
A=a,\alpha\, I\, , 
\end{equation}

\noindent where $a=0,\ldots,9$  and

\begin{equation}
C=   \sum _{n=1}^{6}C^{(2n-2)}\, ,
\hspace{1cm}
C^{(2n-2)} = \frac{1}{(2n-2)!} dZ^{M_{1}}\ldots dZ^{M_{(2n-2)}} 
C^{(2n-2)}{}_{M_{(2n-2)}\ldots M_{1}}\, .
\end{equation}

\noindent Note that the superfield ${\cal B}^{(10)}$ does not occur in
the action (\ref{ad9}). 

The super D9--brane action (\ref{ad9}) has the following symmetries:

\noindent{\bf Worldvolume reparametrizations:}

\begin{equation}
\left\{
\begin{array}{rcl}
\delta_{\eta} Z^{M} & = & \eta^{i} \partial_{i}Z^{M}\, ,\\
& & \\
\delta_{\eta} V_{i} & = & \eta^{j}\partial_{j}V_{i} 
+(\partial_{i}\eta^{j}) V_{j}\, .\\
\end{array}
\right.
\end{equation}

\noindent{\bf $\kappa$-symmetry transformations:}

\begin{equation}
\label{kappa}
\left\{
\begin{array}{rcl}
\delta_{\kappa} Z^{M} E_{M}{}^{a} & = & 0\, ,\\
& & \\
\delta_{\kappa} Z^{M}E_{M}{}^{\alpha I} & = & \left[\bar{\kappa}
\left(1+\Gamma\right) \right]^{\alpha I}\, ,\\
& & \\
\delta_{\kappa} Z^{M} & = &  \left[\bar{\kappa}
\left(1+\Gamma\right) \right]^{\alpha I}E_{\alpha I}{}^{M}\, ,\\
& & \\
\delta_{\kappa} V_{i} & = & E_{i}{}^{A}\delta_{\kappa}E^{B} B_{BA}\, ,\\
\end{array}
\right.
\end{equation}
\noindent where $\Gamma$ is defined by
\begin{equation}
\label{Gamma}
\left\{
\begin{array}{rcl}
\Gamma & = & \frac{\sqrt{|g|}}{\sqrt{|g+{\cal F}|}}  
\sum_{n=0}^{5}\frac{(-1)^{n}}{2^{n}n!} \gamma^{j_{1}k_{1}\ldots j_{n}k_{n}}
{\cal F}_{j_{1}k_{1}}\ldots{\cal F}_{j_{n}k_{n}} 
 {\cal P}_{n} \otimes \Gamma_{(0)} \, ,\\
& &  \\
\Gamma_{(0)} & = &\frac{1}{10!\sqrt{|g|}} \epsilon^{i_{1}\ldots i_{10}}
\gamma_{i_{1}\ldots i_{10}}= {E\over |E|}\Gamma_{11}\, ,
\hspace{.5cm} E = {\rm det}\ E_i{}^a
\hspace{.5cm}
\Gamma_{(0)}^{2}=\unitmatrixDT_{32\times 32}\, ,\\
& & \\
\gamma_{i} & = &E_{i}{}^{a}\Gamma_{a}\, , \\
\end{array}
\right.
\end{equation}
\noindent and satisfies
\begin{equation}
  \Gamma^{2}=\unitmatrixDT_{64\times 64}\, .
\end{equation}

The presence of ${\cal P}_n$ (see eq.~(\ref{Pn})) in the definition of
$\Gamma$ ensures that $\Gamma$ can be written as \cite{Sl2}:

\begin{equation}
\label{gamma}
\Gamma = 
\left(
\begin{array}{cc}
0  & \gamma \\
\tilde{\gamma} & 0 \\
\end{array}
\right)
\end{equation}
with
\begin{equation}
  \gamma\tilde\gamma = \tilde\gamma\gamma =\unitmatrixDT_{32\times 32}\,.
\end{equation}

Note that $\kappa$-symmetry requires that the IIB supergravity fields
satisfy their equations of motion. This does not lead to any restriction
on $C^{(10)}$\ \footnote{Observe that $C^{(10)}$ is a background field.
We are not supposed to take the equation of motion that follows from
varying $C^{(10)}$ in the super D9--brane action. Clearly this equation
of motion would be inconsistent.}. 

\noindent{\bf Target-space super-reparametrizations:}

All superspace fields transform as supertensors under the
super-reparametrization (including general coordinate transformations)

\begin{equation}
\left\{
\begin{array}{rcl}
\delta_{K} Z^{M} & = & - K^{M}(Z)\, ,\\
& & \\
\delta_{K} V_{i} & = & \Delta_{i}\, ,\\
\end{array}
\right.
\end{equation}

\noindent where $\Delta_{i}$ is defined through the supersymmetry 
transformation of $B_{ij}$

\begin{equation}
\delta_{K} {\cal B}_{ij}  =  2\partial_{[i}\Delta_{j]}\, .
\end{equation}

\noindent In a flat background we have

\begin{equation}
  K^\mu = a^\mu - \tfrac{1}{2}\bar\epsilon\Gamma^\mu\theta \,,\qquad
  K^{\dot{\alpha}I} = \epsilon^{\dot{\alpha}I}\,.
\end{equation}

\noindent{\bf $\mathbb{Z}_2$ transformations:}

All $\mathbb{Z}_2$ symmetries of the IIB supergravity background
discussed in the previous Section can be extended to the full super
D9--brane action as follows:

\begin{enumerate}

\item ${\bf \sigma^{3}-symmetries}$: 
The $\sigma^3$--symmetry must be supplemented by 

\begin{equation}
  \theta \rightarrow \pm\sigma^{3}\theta\,,
\end{equation}

\noindent and by a worldvolume parity transformation $\Omega_{\rm D9}$.
We would like to stress the fact that $\theta$ does not change parity
under worldvolume parity transformations because it is a worldvolume
(anticommuting scalar and a target-space spinor).

Note that the action of the $\mathbb{Z}_2$ symmetry on the pulled-back
superfields is in form the same as the action (\ref{s3}) on the
component background fields. 
  
\item ${\bf \sigma^{1}-symmetries}$: The action (\ref{s1}) on the
background fields must be supplemented by the following action on the
worldvolume fields:

\begin{equation}
\left\{
\begin{array}{rcl}
\theta & \rightarrow  &  \pm\sigma^{1}\theta\, ,\\
& & \\
V & \rightarrow & -V \, .\\
\end{array}
\right.
\end{equation}

Again, the action of the pulled-back superfields is in form the same as
the action (\ref{s1}) on the component background fields. 

\end{enumerate}


\section{Truncations of the Super D9--Brane Action}

In the previous Section we have seen that the $\mathbb{Z}_2$ symmetries
of the IIB background can be extended to $\mathbb{Z}_2$ symmetries of
the full super D9--brane action. In this Section we will discuss the
corresponding truncations. In particular, we will investigate what
happens with the fermionic symmetries ($\kappa$-symmetry and
target-space super-reparametrizations) in each case. 

For the explicit results (action and transformation rules) we will limit
ourselves to a flat background. This will lead to two different $N=1$
globally supersymmetric field theories.  Our calculations ensure that
these can be extended to the full $N=1$ local supersymmetry. To compare
we have also included the flat background truncation which leads to an
$N=2$ global supersymmetry. 


\subsection{Flat Background Truncation}

This case has been discussed in \cite{Sl2}.  The flat background
truncation is defined by

\begin{equation}
  e_\mu{}^a  =  \delta_\mu{}^a\, ,\hskip .5truecm ({\rm all\ other\ background\
  fields\  zero})\, .
\end{equation}

The (bosonic and fermionic) embedding coordinates remain nonzero.  There
are two unbroken supersymmetries with {\sl constant} parameter
$\epsilon$. In the appendix we have given the expansion of the
superfields (up to terms linear in $\theta$) corresponding to flat IIB
superspace. 

The Born-Infeld part of the action (\ref{D9BI}) takes the form

\begin{equation}
\label{flatBI}
  {\cal L} = - \sqrt{ |M_{ij}| }\,,
\end{equation}

\noindent where

\begin{equation}
\label{flatM}
  M_{ij} = \eta_{ij} + F(V)_{ij}
  + \bar\theta\Gamma_{(i}\partial_{j)}\theta
  + \bar\theta\sigma^3\Gamma_{[i}\partial_{j]}\theta
  + \tfrac{1}{4}\bar\theta \Gamma^a\partial_{i}\theta\bar\theta
      \Gamma_a\partial_{j}\theta
  + \tfrac{1}{4}\bar\theta \sigma^3\Gamma^a\partial_{[i}\theta\bar\theta
    \Gamma_a\partial_{j]}\theta\,.
\end{equation}

\noindent The Wess-Zumino term (\ref{D9WZ}) retains its form, but now
with flat background RR superfields.

The symmetry transformations are:

\begin{equation}
\label{flatsymm}
\begin{array}{rcl}
  \delta\bar\theta & = & -\bar\epsilon +\bar\kappa(1+\Gamma) + 
          \eta^j \partial_i\bar\theta\,,\\
  &&\\
  \delta x^\mu & = & \frac{1}{2}\bar\epsilon\Gamma^\mu\theta
      +\frac{1}{2}\bar\kappa(1+\Gamma)\Gamma^\mu\theta
      +\eta^i\partial_i x^\mu\,,\\
  &&\\
  \delta V_i & = & -\frac{1}{2}\bar\epsilon\sigma^3
        \Gamma_\mu\theta \partial_i x^\mu
    -\frac{1}{24}(\bar\epsilon\sigma^3\Gamma^a\theta
                  \bar\theta\Gamma_a\partial_i\theta
       + \bar\epsilon\Gamma^a\theta
                  \bar\theta\sigma^3\Gamma_a\partial_i\theta) \\
    &&\\
    &&
  -\frac{1}{2}\bar\kappa(1+\Gamma)\Gamma_\mu\sigma^3\theta\partial_i x^\mu
  -\frac{1}{8}\bar\kappa(1+\Gamma)\Gamma^a\theta
   \bar\theta\Gamma_a\sigma^3\theta \\
    &&\\
    && -\frac{1}{8}\bar\kappa(1+\Gamma)\Gamma^a\sigma^3\theta
   \bar\theta\Gamma_a\theta \\
    &&\\
    && + \eta^j\partial_j V_i + (\partial_i \eta^j) V_j\,.
\end{array}
\end{equation}

Note that the flat background truncation preserves the
$\kappa$-transformations. At this stage these $\kappa$-transformations
can be gauge-fixed as described in \cite{Sl2}. We will see that the
Heterotic and Type I truncation correspond, in a flat background, to
truncations of (\ref{flatBI}) and (\ref{flatsymm}) in which the
$\kappa$-transformations are automatically eliminated.  These
truncations are described below. 

\subsection{Heterotic Truncation}

The Heterotic truncation is defined by the discrete symmetry of the
D9-brane generated by $\pm\sigma^{3}$:

\begin{equation}
C^{(2n-2)}=(1\mp\sigma^{3})f=0\, ,
\hspace{1cm}
n=1,\ldots,6\, .
\end{equation}

\noindent The $N=1$ Heterotic superspace is defined by the truncation

\begin{equation}
(1\mp\sigma^{3})\theta = 0\, .  
\end{equation}

\noindent The action after truncation is given by

\begin{equation}
\label{HetBI}
S_{\rm Heterotic} = -\int_{M^{10}} d^{10}\xi e^{-\varphi} 
\sqrt{|g_{ij}+{\cal F}_{ij}|}\, ,
\end{equation}

\noindent where it is understood that all superfields are $N=1$
Heterotic, i.e., $\sigma^3$--truncated IIB superfields.  On-shell they
describe the Heterotic supergravity multiplet (\ref{heterotic}).  The
Wess-Zumino term vanishes in this truncation. 

We now take a flat background and choose the sign for the $\sigma^3$
truncation such that $\theta_2=0$ \footnote{ The other sign leads to the
same result with $\theta_1 =0$.}.  Then the truncation of the IIB
background fields implies that $\epsilon_2=0$. The remaining symmetry
parameters have to be constrained by the condition $\delta\theta_2=0$,
which means that $\bar\kappa_2+\bar\kappa_1\gamma=0$. This implies that
$\bar\kappa(1+\Gamma)$ vanishes. At this stage it is convenient to go to
the static gauge, i.e., to choose $x^\mu = \delta^\mu_i\xi^i$.  This
requires a compensating world-volume reparametrization with parameter

\begin{equation}
\label{eta}
  \eta^\mu = -\tfrac{1}{2}\bar\epsilon_1\Gamma^\mu\theta_1\,.
\end{equation}

The result of this truncation in a flat background is given by
an  action of the form (\ref{flatBI}), now with

\begin{equation}
\label{HetM}
  M_{\mu\nu} = \eta_{\mu\nu} + F(V)_{\mu\nu}
  + \bar\chi\Gamma_{\mu}\partial_{\nu}\chi
  + \tfrac{1}{4}\bar\chi \Gamma^a\partial_{\mu}\chi\bar\chi
      \Gamma_a\partial_{\nu}\chi\,,
\end{equation}

\noindent where we have set $\chi\equiv\theta_1$.  This action is
invariant under the supersymmetry transformations
($\epsilon\equiv\epsilon_1$)

\begin{equation}
\label{Hetsusy}
\begin{array}{rcl}
  \delta\bar\chi & = & -\bar\epsilon
    + \eta^\mu\partial_\mu\bar\chi\,, \\
  &&\\
  \delta V_\mu & = & -\frac{1}{2}\bar\epsilon\Gamma_\mu\chi
    - \frac{1}{12}\bar\epsilon
      \Gamma^a\chi\bar\chi\Gamma_a\partial_\mu\chi
    + \eta^\rho\partial_\rho V_\mu + (\partial_\mu\eta^\rho)V_\rho\,,
\end{array}
\end{equation}

\noindent with $\eta$ given by (\ref{eta}).

Note that the
 Heterotic $\mathbb{Z}_2$ also involves a world volume parity
transformation $\Omega_{D9}$.  This acts only on the Wess-Zumino term,
and in the form of $\Gamma$ (see the $\kappa$-transformation rules).
However, the truncation sets the Wess-Zumino term equal to zero, and
eliminates the $\kappa$-transformations, as we have seen above.
Therefore the world volume parity transformation becomes irrelevant.
This is important in going to the static gauge, which turns the
D9-theory into an ordinary $D=10$ field theory. Before truncation the
static gauge would have been inconsistent with $\Omega_{D9}$, since this
gauge choice identifies $\Omega_{D9}$ with the target space parity
transformation, which is not a symmetry of the IIB multiplet. Therefore
taking the static gauge before truncation would break the $\mathbb{Z}_2$
symmetry of the D9-brane, while after truncation there is no problem. 

The Heterotic truncation implies that a locally (nonlinear)
supersymmetric extension of the $D=10$ Maxwell theory exists which
contains a cosmological constant\footnote{Since there is no
  superalgebra including the bosonic group $SO(2,10)$ this statement
  needs some explanation. If the Einstein $R$ term is added to the
  action, the equations of motion of the theory allow for a domain
  wall solution whose metric, in an appropriate ``dual'' frame, is
  identical to $AdS_{10}$ spacetime and therefore has the isometry
  group $SO(2,10)$. The same happens for the standard D8-brane
  solution \cite{kn:PW,kn:BRGPT}. However, as opposed to the D3-brane,
  the solution has a nontrivial dilaton (this is the dilaton that
  multiplies the cosmological constant) that which is not invariant
  under the $AdS_{10}$ isometries which are not symmetries of the full
  background.}.  As far as we know the existence of such an $N=1$
supersymmetrization has not been noted before in the literature.


\subsection{Type I Truncation}

The Type I truncation is associated to the discrete symmetry of the
D9-brane generated by $\pm\sigma^{1}$:

\begin{equation}
\label{VA}
{\cal B} =C^{(2n-2)}=(1\pm\sigma^{1})f=V_{i}=0\, ,
\hspace{1cm}
n=1,3,5\, .
\end{equation}

\noindent The Type I superspace is defined by the truncation

\begin{equation}
(1\pm\sigma^{1})\theta =  0\, .  
\end{equation}

\noindent The action after truncation is given by

\begin{equation}
\label{VAaction}
S_{\rm Type\ I} = -\int_{M^{10}} d^{10}\xi \left\{ e^{-\varphi} 
\sqrt{|g_{ij}|} +  C^{(10)}\right\}\, .
\end{equation}

\noindent It is understood that all superfields are $N=1$ Type I, i.e.,
$\sigma^1$--truncated IIB superfields.  On-shell they describe the Type
I supergravity multiplet (\ref{typeI}). 

We now consider a flat background.  Note that in this truncation we have
set $V_i=0$\footnote{ The elimination of the BI vector is natural from
the string theory point of view where orientifolding the Type IIB
superstring with $\Omega$ changes the Chan-Paton factors from U(N) to
SO(N). In our case, for a single D9, we have $N=1$ and hence no BI
vector field. We thank B. Schellekens for pointing this out to us.}.
This simplifies the $\kappa$-transformations considerably, since now
$\Gamma = \sigma^1\otimes \Gamma^{(0)}$. In particular, in the flat
background which we will consider below $\Gamma^{(0)}=\Gamma_{11}$.
Therefore

\begin{equation}
  \bar\kappa(1+\Gamma) \to \bar\kappa
 \left(
\begin{array}{cc}
  1  & -1 \\
 -1  &  1 \\
\end{array}
\right)\, .
\end{equation}

Let us choose the truncation $\theta_1-\theta_2=0$ \footnote{The other
truncation, $\theta_1 + \theta_2 = 0$, leads to a surviving
kappa-symmetry.  The action, however, in this truncation becomes zero
(see below).}, and set $\chi=\theta_1/\sqrt{2}$. The truncation of the
IIB background requires also $\epsilon = \epsilon_1/\sqrt{2} =
\epsilon_2/\sqrt{2}$.  The remaining supersymmetry transformations are,
in the static gauge and in a flat background

\begin{equation}
\label{VAsusy}
  \delta\bar\chi  =  -\bar\epsilon
    + \eta^\mu\partial_\mu\bar\chi\,, 
\end{equation}

\noindent with $\eta$ as in (\ref{eta}). This leaves invariant the
action (\ref{VAaction}), with $\varphi=0$ and

\begin{equation}
\label{VAM}
  g_{ij}\to g_{\mu\nu} = \eta_{\mu\nu} 
  + \bar\chi\Gamma_{\mu}\partial_{\nu}\chi
  + \tfrac{1}{4}\bar\chi \Gamma^a\partial_{\mu}\chi\bar\chi
      \Gamma_a\partial_{\nu}\chi\, .
\end{equation}

This action is the Volkov-Akulov action \cite{kn:VA} generalized to ten
dimensions\footnote{The occurrence of a Volkov-Akulov action, after
truncation of the BI vector was also mentioned in \cite{kn:Ren}.
Volkov-Akulov actions also occur in truncated and gauge-fixed versions
of non-spacetime-filling branes \cite{ap}.}.

We observe that the Wess-Zumino term in the original D9-brane action is
separately invariant under supersymmetry transformations: only
$\kappa$-invariance, which in this truncation disappears, requires a
collaboration of the Born-Infeld and the Wess-Zumino term.  Therefore
the Born-Infeld and the Wess-Zumino term should be separately invariant
under the transformation (\ref{VAsusy}).  The existence of two
Volkov-Akulov invariants appears unlikely, and indeed, a closer look at
the expansion in $\chi$ shows that, in a flat background, they are equal
up to an additive constant\footnote{Note that in the other truncation,
$\theta_1 + \theta_2 = 0$, $C^{(10)} \rightarrow - C^{(10)}$ and the
kinetic and WZ term cancel against each other.}:

\begin{equation}
\sqrt {|g_{\mu\nu}|} - \sqrt{|\eta_{\mu\nu}|} = C^{(10)}\, .
\end{equation}
 
Thus the Wess-Zumino term, expressed in $C^{(10)}$ only, provides an
interesting reformulation of the Volkov-Akulov action.  This Wess-Zumino
construction of a Volkov-Akulov theory can be done in any dimension
which allows spacetime filling $D$-branes, i.e., $D=3,4,6$ and $10$. 
 

\section{Conclusions}

In this letter we have discussed two inequivalent truncations of the
super D9-brane. Only one of them, the Type I truncation, has a natural
string theory origin. It is the truncation that is triggered by the
worldsheet parity operator $\Omega$ of the fundamental string. We only
considered the truncation of a single super D9-brane and this leads to a
$D=10$ Volkov-Akulov action in terms of a single Majorana-Weyl fermion. 

As a by-product we found the following interesting reformulation of the
$N=1, D=10$ Volkov-Akulov action\footnote{ Similar results hold in any
spacetime dimension in which spacetime-filling D-branes exist, i.e., D=
3,4,6, and 10.}:

\begin{equation}
S_{\rm Volkov-Akulov} = \int_{M^{10}} d^{10} x \  C^{(10)}\, ,
\end{equation}

\noindent where $C^{(10)}$ is the pull-back of a 10-form superfield in
the Type I truncated flat superspace. Its 11-form curvature $G^{(11)} =
d C^{(10)}$ has non-zero components given by

\begin{equation}
G^{(11)}_{\alpha I \beta J a_1\cdots a_n} = \tfrac{i}{2} 
\left (\Gamma_{a_1 \cdots a_n}(1+\Gamma_{11})\right )_{\alpha\beta} 
(\sigma^1)_{IJ}\, .
\end{equation}

\noindent Such superfields have explored in the past \cite{gates} and
here we find a nice application of them. 

It would be interesting to extend our work to the non-Abelian case and
make contact with the supersymmetric low-energy effective action of the
Type I $SO(32)$ superstring. The (truncated version of the) worldvolume
action of 32 coincident D9-branes should be supplemented with a
contribution from a single orientifold O9-plane to give the Type I
effective action:

\begin{equation}
S_{\rm 32\ D9} + S_{\rm O9} = S_{IO_{32}}\, .
\end{equation}

Some of the coupling terms in $S_{\rm O9}$ have been calculated
\cite{Mukhi}, for more recent results, see \cite{kn:Ser}.  Apart from
this we need to find a kappa-symmetric non-Abelian generalization of the
super D9 brane action. It would be very interesting to find such a
generalization.


\section*{Acknowledgements}

We would like to thank A. Ach\'ucarro, C.~Bachas and B.~Schellekens for
useful discussions.  This work is supported by the European Commission
TMR program ERBFMRX-CT96-0045, in which E.B.~and M.d.R.~are associated
to the University of Utrecht. E.B. wishes to thank the institute for
Theoretical Physics at the Universidad Autonoma de Madrid, where part of
this work was done, for its hospitality.  The work of T.O.~and B.J.~has
been supported in part by the by the European Union TMR program
FMRX-CT96-0012 {\sl Integrability, Non-perturbative Effects, and
Symmetry in Quantum Field Theory}. The work of T.O.~has also been
suported by the Spanish grant AEN96-1655.  T.O.~would also like to thank
the University of Groningen for its warm hospitality during the early
stages of this work. 

\appendix

\section{Expansion of flat Superfields}

In this Appendix we collect the value of the relevant superfields in a
flat background. 

The target-space super-reparametrization parameter $K^M$ is determined
by

\begin{equation}
K^{\mu}  =  a^{\mu} - \tfrac{1}{2}
   \bar{\epsilon}\Gamma^{\mu}\theta \,,\qquad
K^{\dot{\alpha}I}  =  \epsilon^{\dot{\alpha} I} \, .
\end{equation}

\noindent The supervielbein $E_M{}^A$ takes on the form

\begin{equation}
\begin{array}{rcl}
E_{\mu}{}^{a} & = & \delta_{\mu}{}^{a}\, , \hskip 1.5truecm
E_{\dot{\alpha} I}{}^{a}  =  
- \frac{1}{2}\left(\bar{\theta} \Gamma^{a}\right)_{\dot{\alpha} I}\, ,\\
&&\\
E_{\mu}{}^{\alpha I} & = & 0\, , \hskip 1.8truecm
E_{\dot{\alpha} I}{}^{\alpha J}=\delta_{\dot{\alpha} I}{}^{\alpha J}\, .
\end{array}
\end{equation}

\noindent The superfields ${\cal B}$ and $C^{(2n-2)}$ are more
complicated.  For our purposes we only need the terms linear in $\theta$
which are given by

\begin{equation}
\begin{array}{rcl}
{\cal B}_{\mu\dot{\alpha} I} & = & -\frac{1}{2}\left(\bar{\theta} \sigma^{3}
 \Gamma_{\mu}\right)_{\dot{\alpha} I}
\, . \\
&&\\
C^{(2n-2)}{}_{\mu_{1}\ldots\mu_{2n-3}\dot{\alpha} I} & = & 
 \frac{1}{2}
 \left(\bar\theta{\cal P}_{n}\Gamma_{\mu_{1}\ldots\mu_{2n-3}}
  \right)_{\dot{\alpha} I}
\, .\\
\end{array}
\end{equation}

%
%
%
%
%
%
%
%
%
%
%
%

\end{document}